\documentclass[10pt]{article}
\usepackage{graphicx,amssymb, amstext, amsmath, epstopdf, booktabs, verbatim, gensymb, geometry, appendix, lmodern,bchart,pgfplots,url}
\usepackage{chngcntr}
 
\counterwithout{figure}{section}
\counterwithout{table}{section}

\pgfplotsset{width=7cm,compat=1.8}

\newcommand*\Title{KOALA Project Report 2}
\newcommand*\cpiType{--- KOALA Project Report 2}
\newcommand*\Date{April 2018}

\title{Are Children Well-Supported by Their Parents Concerning Online Privacy Risks, and Who Supports the Parents?}

\author{Jun Zhao}
\date{\today}
\usepackage{cpi} % This is what makes your document look like a cpi document.

\begin{document}

\begin{titlepage}
\maketitle
\end{titlepage}

\linespread{1.15} %Set standard document linespacing

\begin{executive}

Tablet computers are becoming ubiquitously available at home or school for young children to complement education or entertainment. However, parents of children aged 6-11 often believe that children are too young to face or comprehend online privacy issues, and often take a protective approach to restrict or monitor what children can access online, instead of discussing privacy issues with children. Parents work hard to protect their children's online safety. However, little is known how much parents are aware of the risks associated with the implicit personal data collection by the first- or third-party companies behind the mobile `apps' used by their children, and hence how well parents can safeguard their children from this kind of risks.

Parents have always been playing a pivotal role in mitigating children's interactions with digital technologies --- from TV to game consoles, to personal computers --- but the rapidly changing technologies are posing challenges for parents to keep up with. There is a pressing need to understand \textit{how much parents are aware of privacy risks concerning the use of tablets and how they are managing them for their primary school-aged young children}. At the same time, we must also reach out to the children themselves, who are on the frontline of these technologies, to learn \textit{how capable they are to recognise risks and how well they are supported by their parents to cope with these risks}. Therefore, in the summer of 2017, we conducted face-to-face interviews with 12 families in Oxfordshire and an online survey with 250 parents. This report summarises our key findings of these two studies.

\frame{
\textbf{Three key findings}
\begin{enumerate}
\item Parents commonly associated privacy risks with access to the Internet, exposure to inappropriate content, in-app adverts, or strangers. However, \textbf{their knowledge of personal data collected by mobile apps is low}. 

\item Parents often think their children are too young to understand privacy risks online and delay these conversations with their children. As a result, children often rely on their parents' guidance to cope with unknown risks, and \textbf{they not always capable of recognising personal privacy-related risks}.

\item When children do seek help from their parents, parents do not necessarily fully understand the risks themselves. Existing privacy safeguarding technologies mainly focus on enabling content control and \textbf{offer little choices for raising parents' awareness of personal data collection risks or supporting their children's learning}.

\end{enumerate}
}

These findings provide critical input for us to consider providing better support for parents to scaffold their children's ability of recognising and coping with online privacy risks, particularly those associated with implicit personal data tracking. 

\end{executive}

\section{Introduction}
% population and wide adoption of tablets
Tablet computers are widely used to complement education and provide entertainment at home and school. In the UK, 44\% of children aged between 5 to 11 have their own tablet computers or use them as the major means to go online~\cite{ofcom2016}. Tablet computers are used in over 70\% of UK schools to complement education. A similar trend is also observed in the US, where ownership of tablets by young children has grown 5 times between 2011 and 2013 \cite{commonsense2013}. This is transforming the learning experience for young children~\cite{judge2015using,papadakis2016comparing,neumann2014touch,schacter2016improving}.

%tablets pose new privacy risks to children % families and children struggle to cope with this
However, these new ways of online access pose new privacy risks to families and young children. The large number of applications (`apps') that can be downloaded for free are a major way in which children interact with these devices. Currently these `free' apps are largely supported by monetisation of user's personal information~\cite{acquisti2016economics,kummer2016private}. While the use of cute characters may suggest benign practices, a large amount of personal information and online behaviour may be collected from children's apps and shared with third party online marketing and advertising industry~\cite{reyes2017our}. Such practices are not unique to children's apps, but young children are particularly vulnerable to this type of exposure and less capable to resist the resulting (personalised) in-app advertisements and game promotions~\cite{digital2017}. 

% existing studies on technologies and families show a mediation process is most effective
%The need for child-specific privacy protection is recognised, as seen by the launch of kids app stores, device-specific parental restriction mechanisms, and children-specific regulations (e.g. the Children's Online Privacy Protection Act or COPPA). However, unsolicited personal data collections are still identified in apps designed for children~\cite{federal2015ftc,reyes2017our}, and parents and children lack an easy way to learn about their data privacy risks. 

Previous work~\cite{zhang2016nosy,kumar2018cscw} have discussed different types of online privacy threats that are recognisable by young children interacting with mobile devices, but personal data collection by mobile apps has not been specifically discussed. Exiting research has shown that this opaque data privacy risk is unknown to most users and extremely challenging to manage~\cite{van2017better,vallina2016tracking}. Current internet safeguarding advice for young children mainly focuses on the management of content, cyberbullying or social media~\cite{digital2017}. We argue that approaches to raising young children's awareness of data privacy risk are under-researched, yet are critical for scaffolding children's mental models of privacy. 

Our work thus focuses on understanding the current practices applied by parents for safeguarding the online privacy of their children, and identifying barriers faced by families for mediating data privacy risks effectively. We aim to explore three questions: 
\begin{itemize}
\item What privacy risks are children aged 6-10 aware of when interacting with their tablets?
\item How much are parents aware of the possibility of personal data collection by the apps regularly used by their children?
\item What technical and social practices are used by parents for safeguarding their young children's online privacy, and what support do they need to mediate data privacy risks?
\end{itemize}

To achieve this, we conducted semi-structured interviews with 12 parent-child pairs from the South East area of the UK, complemented by an online survey with 220 parents from western developed countries (78.6\% UK residence). Our analysis shows that current privacy tools on tablet devices provide good support for parents in our study to set up basic content control and filtering, but offer only limited capabilities for supporting them to gain awareness of data privacy risks or help their children establish this understanding. Although savvy tablet users, children in our study did not demonstrate a consistent understanding of their risk coping actions. Similarly, their parents -- who are largely relied upon by young children -- had limited awareness of the implicit personal data collection by third-parties or how to safeguard their children from this kind of risks. Our findings suggest that we are in need of new tools to raise parents' awareness of the current personal data collection practices of mobile applications, and help both parents and young children to gain the skills for coping with data privacy threats on mobile platforms. 

%To alleviate their concerns, parents largely rely on applying an arms-length approach to their young children, e.g. setting up technical restrictions or establishing specific social practices at home. Although these approaches provide certain protections, they do not necessarily help young children understand the actual privacy risks. As a result, young children have critical gaps in their privacy knowledge and largely rely on parents' help; and parents felt challenging to keep informed. 

\section{Background}

%The gap between stated privacy concerns and actions can be due to the lack of self-awareness~\cite{almuhimedi2015your} or lack of  options~\cite{brandimarte2013misplaced}. Privacy decision making is a complex process, and we are still seeking better ways to balance the complexity of information and an individual's autonomy. In this work, we focus on understanding parents' and young children's current awareness of personal data privacy threats so that we can better facilitate their needs. \comm{UL: i think this entire paragraph can be skipped}

\subsection{Children and Mobile Technology}
In 2016, UK children aged 5-15 were for the first time reported to spend more time online than watching TV ~\cite{ofcom2016}.  Tablets and smartphones are becoming the most popular devices used by children to go online, overtaking laptops or computers. However, little is known about the impact of these technologies on children's online safety and general well-being~\cite{digital2017}.

Research has shown that teenagers and tweens have a reasonable understanding about privacy risks online and their implications, and strategies to protect themselves~\cite{davis2013tweens,mostmans2014would}, although they are facing increasing challenges related to the use of Social Media platforms and smartphone devices~\cite{wisniewski2017parental}. 
However, younger children are generally expected to be less capable to engage with the Internet in a safe and mindful way~\cite{shmueli2010privacy,holloway2013zero}. Using hypothetical scenarios and the contextual integrity framework, Kumar et al.~\cite{kumar2018cscw} unpacked young children's (aged 5-11) understandings of online privacy and identified key gaps in their recognition of how information is transmitted from/to their devices. Zhang-Kenney et al.~\cite{zhang2016nosy} found that children aged 7-11 perceived privacy related to their use of mobile devices as being alone, hiding secrets, keeping things to yourself, or not talking to strangers, largely drawing on their offline life experiences.  We extend this existing body of work and examine children's understanding about data privacy on the mobile platforms in a contextualised and nuanced way. 

\subsection{Parental Mediation of Mobile Technology}
%
 %Research has shown that a combination of technical restrictions with active discussions or co-engagement can be more effective for reducing the access to age-inappropriate content or adverts~\cite{chen2013app,holloway2013zero,hashish2014involving,ofcom2016}  parents do find that they are predominantly struggling with controlling screen time~\cite{hiniker2016not,hiniker2016screen}, . Our work contributes to understandings about privacy concerns of both parents and children (aged 6-10) in relation to interaction with tablet computers, and the alignment between parents' expectations of their children's awareness of privacy risks and the actual privacy mental model of their children~\cite{slovak2016scaffolding,hashish2014involving}.
%
Researchers have found that parents are generally concerned about their children's online privacy risks~\cite{ofcom2016,zhang2016nosy}, and employ a range of mechanisms in an attempt to safeguard their children, such as password protections, supervising them, or setting up social practices at home (e.g. limiting access to the Internet)~\cite{ofcom2016}. 
Our work aims to investigate whether UK parents of young children are effectively supported to choose less privacy-invasive mobile apps for their children and what parents may struggle with. %Previous work~\cite{zhang2016nosy,kumar2018cscw} have explored parents' general awareness of online privacy and security risks related to their children's usage of mobile devices. However, concerns about personal data collection by mobile apps were not explicitly discussed in these studies, and parents' and children's awareness of and coping strategies with these risks are little known. 

%making an informed decision is a complex task~\cite{egelman2013choice,liccardi2014no,van2017better}. 
%Even for grown-ups, making informed privacy choices is a very challenging task~\cite{egelman2013choice,liccardi2014no,van2017better}. %We are interested to find out whether young children are capable of acting on new privacy risk scenarios using existing knowledge scaffolding acquired from family social practices, educations at schools, or influences from friends or siblings~\cite{}. 

%\subsection{Parental Mediation with Mobile Technologies}
A large body of previous research has investigated parents' role in regulating their children's' interactions with technologies, such as TV or computers~\cite{lin1989parental,austin1993exploring,valkenburg1999developing,lwin2008protecting,valcke2010internet,shin2011parental}. Parents' role in mediating children's usage of the new mobile technologies are largely investigated for parents of older children~\cite{wisniewski2017parental,wisniewski2017parents} or parents battling with young children's screentime control~\cite{hiniker2016screen,hiniker2016not} or content filtering~\cite{hashish2014involving}. These studies highlighted the challenges that parents face with fully grasping the new technologies or socially engaging their young children. 

Literature suggests that parents' engagement largely falls into three patterns~\cite{hiniker2016not}: \textit{active mediation}, where parents take time to discuss with children about usage of technologies~\cite{livingstone2008parental}; \textit{restrictive mediation}, where access to technologies is controlled and limited; and \textit{co-engagement}, where parents consume technologies together with their children without injecting any critiques~\cite{hashish2014involving}. 
Existing parental mediation approaches for managing older children's online privacy and safety have largely focused on a \textit{restriction} approach, for example, controlling or monitoring teens' interaction with their mobile phones using parental control software. However, the usability of these tools and their effectiveness has been questioned~\cite{livingstone2008parental,anderson2016parents,wisniewski2017parental,wisniewski2017parents}. For managing younger children's use of mobile technologies, new solutions are developed to promote active mediation and co-engagement between parents and young children, for example, for setting up content filtering~\cite{hashish2014involving} or limiting smartphone use~\cite{ko2015familync}. Results show that a combination of technical and social mediation can lead to a more productive learning experience and positive parent-child relationship. We extend this literature by investigating approaches currently taken by the parents to manage online privacy risks for their young children (under 11) and their effectiveness.

\subsection{Personal Data Collection on Mobile Platforms}
More broadly, concerns about the collection and use of personal information have been discussed prior to the emergency of the web and smartphones. Westin~\cite{westin1968privacy} describes privacy as the notion of personal autonomy over what information is communicated, by whom, and when. Nissenbaum~\cite{nissenbaum2004privacy} views privacy as a dynamic, dialectical process of boundary negotiation, and argues that there are no universal norms: privacy preferences ought to differ according to distinct culture or context.

Major smartphone operating systems like Android and iOS have endeavoured to improve on their existing privacy permission settings with the introduction of more granular controls. Despite such improvements, studies suggest that existing mechanisms are still inadequate, failing to  raise users' awareness of these settings or convey implications to their privacy risks~\cite{balebako2014privacy,Liccardi2013,van2017better}. In response to these problems, researchers have explored alternative, novel presentations of privacy notifications, by raising users' awareness on the collection of sensitive personal information~\cite{almuhimedi2015your}, making complex privacy notices easier to digest~\cite{kelley2009nutrition,liccardi2014no}, or easing the process of decision by predicting personal privacy preferences~\cite{lin2012expectation,liu2016follow}. More recently, we have seen a growing body of work revealing third-party data collection practices to the users in various ways, such as in an augmented permission interface~\cite{balebako2014privacy} or through alternative visualisations~\cite{van2017better}. The goal of these studies is to reveal the opaque personal data flows, from the mobile apps to first-party app developers as well as third-party online advertising and marketing trackers, and investigate the impact on users' privacy mental model~\cite{vallina2016tracking,yu2016tracking,binns2018toit}. Our study explores the effectiveness of current privacy setting mechanisms for helping parents safeguard their children, and parents' possible needs for the alternative, transparent presentation of personal data risks. 

\section{Methodology}
We undertook a mixed method study, including qualitative semi-structured interviews and an online survey. The former allowed us to look into the practices and understandings of individual child and parent, and the latter complemented these thick descriptions with quantitative data from a larger sample. Our study was reviewed and approved by our university's research ethics committee.

% * <p.slovak@ucl.ac.uk> 2018-01-10T11:36:40.975Z:
% 
% I'd swap the order of the sections to match the temporal ordering? 
% 
% ^.
\subsection{Semi-structured Interviews to Parents and Children}
The parent-child interviews were designed to last no longer than one hour, and consisted of three parts. For safeguarding reasons, parents and children participants of the study stayed in the same room throughout each interview. We acknowledge that the presence of parents in the interviews may impact on the responses provided by the children.

In the first part, each child participant was asked to show us two of their favourite apps or games on their tablet, which was brought along to the study. They were also asked to explain how the apps were found, e.g. by themselves or together with the parents. In this way we were able to look into both the technical and social safeguarding practice at home from the child's point of view. The parent was then asked to comment on their privacy concerns of these favourite apps. 

In the second part of the study, we used three hypothetical scenarios to explore how children felt about personal data collection by the apps, and how they might cope with them. In the three scenarios the child was asked to imagine that an app was asking for access to their camera, microphone or location information and how they felt about it. We then showed each child how to check access permissions on their device, and asked how they felt about their favorite apps having access to their e.g. camera or location. At the end of this session, parents were asked to comment on how they felt about their children's responses.

\begin{figure*}[ht!]
  \centering
  \setlength\fboxsep{0pt}
  \setlength\fboxrule{0pt}
  \fbox{\includegraphics[width=5.2in]{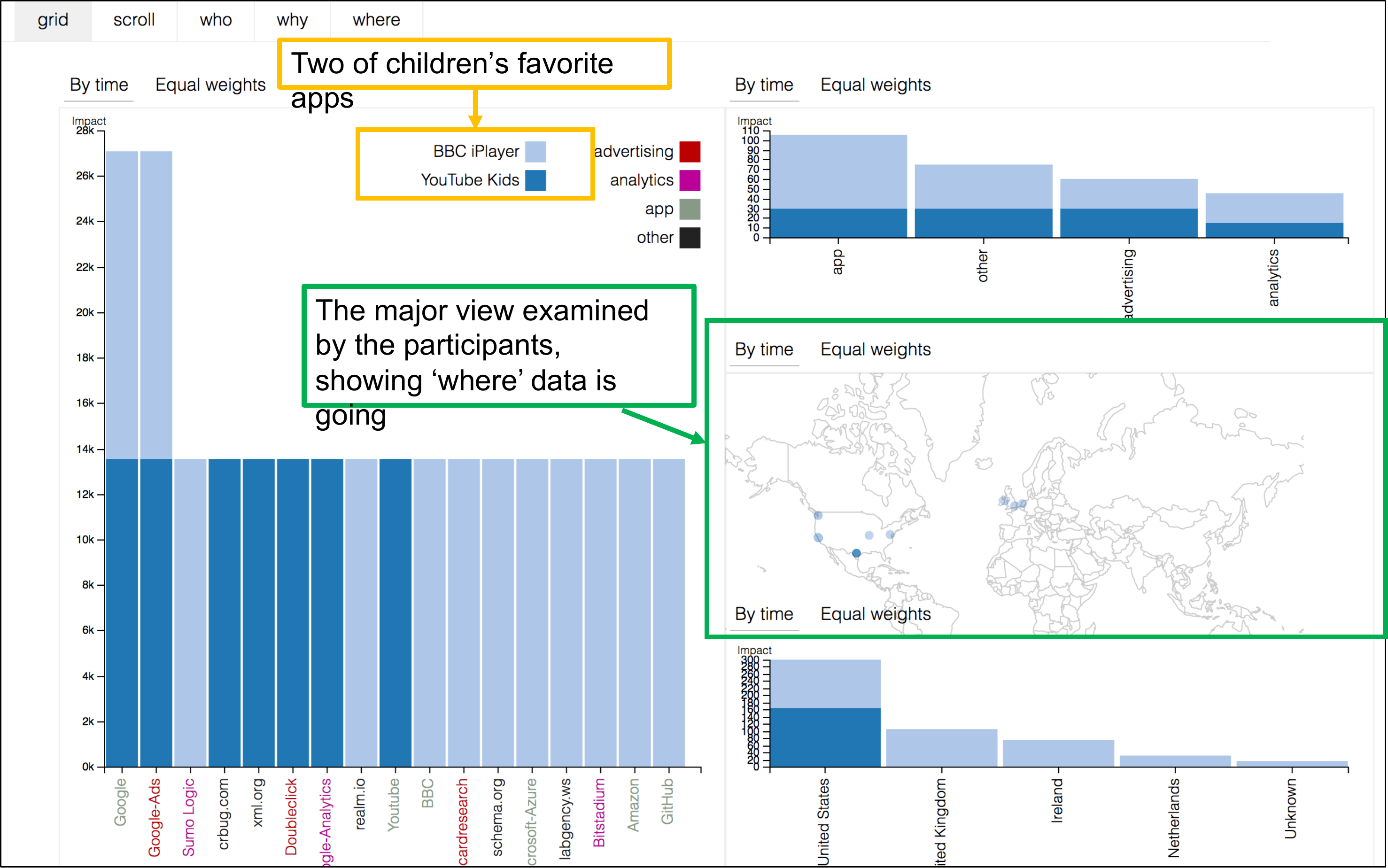}}
  % \fbox{\includegraphics[width=7.5in]{refine_procedure}}

  \caption{The web application used in the last phase of the lab study shows destinations to which personal data may be sent through each app, in a stacked bar chart view (left) and a map view (right). Hovering over an app (in the legend) highlights data trackers specific to that app.}
  \label{fig:tracker}
  \end{figure*}

In the final part of the interview, we showed parents and children a new way of examining possible personal data leakage of their favorite apps using a prototype application from our prior work~\cite{vankleek2018chi} (see Figure~\ref{fig:tracker}). Although this platform was not specifically designed for children, we wanted to see how the explicit presentation of information transmission of mobile apps in this tool would impact participating children and parents' mindset about data privacy risks and their existing safeguarding practices. After giving each child five minutes to explore the interface by themselves, we asked them how they felt about their data being collected by various companies. We concluded the interview by asking some open questions to parents concerning their opinions about current technology and legal support for safeguarding their children's privacy online.  

\subsection{Online Surveys to Parents}
To confirm the themes observed from the 12 families in the lab study, we designed an online survey to parents of young children who regularly interact with mobile devices. The online survey was supported by the Prolific Academic system (\url{http://prolific.ac}),  a crowdsourced online survey platform that is capable of recruiting participants fitting specific criteria. In our survey, participants were required to be parents or guardians to at least one child aged between 6 and 10, who has regular access to a tablet computer or smartphone. The survey was set up to be completed within 15 minutes, and each participant was rewarded \textsterling1.4 for completing the survey. Prolific provides comprehensive demographic information about the participants, such as their levels of family income or residential areas, which enabled us to collect data from a more diverse population than was the case in our lab study.  
% * <p.slovak@ucl.ac.uk> 2018-01-10T11:35:51.700Z:
% 
% >  which is a spin-off of Oxford University
% why is this important? 
% 
% ^.

The online survey contained 3 parts. It started with some basic demographic questions about the parent and child, such as their age, level of education and usage of devices. The parent was then asked to choose one child aged 6-10 to complete the main survey. 

There are two themes in the second/main part of the study: parents' privacy concerns in relation to their children's use of tablets or smartphones, and their knowledge of existing privacy control technologies, including password controls, kids app stores, and app privacy permissions. It started with \textit{generic concerns} considered by parents when choosing apps for their children and then focused on their opinions to specific personal data collection behaviours, such as accessing cameras, location information or sensitive personal information. 

Finally the survey was concluded with a reflection section, asking whether the parents felt they learnt anything new about safeguarding their children's privacy and whether they would communicate this new knowledge to their child. 

\subsection{Participants}
Our lab study participants were recruited through social media, public posters, mailing lists, and snowball sampling. All interviews were conducted in person, and parents and children stayed in the same room throughout. The interview protocol was designed to take 30-45 minutes. Each participant child received a \textsterling15 Amazon Gift Voucher as a token of appreciation. In total we had 12 families and 14 children, with at least one child from each family aged between 6 and 10 (inclusive). Mean age of the children was 7.7. More mothers (10 participants) than fathers accompanied the child to the study.  Siblings were included in the same interview if they were both within the age of 6-10.

% \begin{table}[ht!]
%     \centering
%     \begin{tabular}{c|c|c}
%      Family & Child participants, Age & Device \\ \hline
%       1  & C1 (Boy;10) &Family smartphone\\
%       2  & C2 (Girl;7) & Own Android tablet \\
%       3  & C3 (Boy;10) C4 (Girl;7) & Own smartphone \\
%       4  & C5 (Boy;8) C6 (Boy;6) & Shared iPad \\
%       5 & C7 (Girl;8) & Shared iPad \\
%       6 & C8 (Boy; 8) & Own iPad \\
%       7 & C9 (Boy;10) & Own Kindle Fire \\
%       8 & C10 (Boy;7) & Parent's iPad \\
%       9 & C11 (Boy;6) & Own Kindle Fire\\
%       10 & C12 (Girl;6) &  Own iPad  \\
%       11 & C13 (Girl;8) & Shared family ipad \\
%       12 & C14 (Boy;8) & Own ipad  \\
%     \end{tabular}
%     \caption{Child Participants and Devices Used at Home}
%     \label{tab:my_label}
% \end{table}

 %Altogether we collected 10.5 hours interview audio over August and September 2017. We present our results in four high-level themes: general parental control practices, parents' and children's mental model of privacy, and children's coping strategies.  

The online survey received 250 responses. 29 responses were excluded because their children were outside the age range of 6 and 10. Of the remaining 221 responses, the average age of the parents was 35.9, and 69.5\% were female. Most of our respondents (78\%) were UK residents, 15\% US residents, and the remaining 7\% reside in other western countries. The UK residents were evenly distributed in different regions of the country. The respondents had on average 2.27 children (range 1-5, SD = 0.88), and the average age of the child selected for completing the survey was 7.91 (range 6-10, SD = 1.41).

\subsection{Data Analysis Method}
We collected 10.5 hours of interview audio in total. The audio files were carefully transcribed and Thematic Analysis~\cite{braun2013successful} was applied by 2 researchers to identify common themes. Thematic Analysis allows us to take a ground-up approach to data analysis, without being confined to a prior theoretical framework. The two researchers started by coding half of the transcripts independently. We then discussed the codes that emerged, consolidated them, and organised them into 5 themes: children's understanding, children's coping strategies, parents' concerns, parental mediation methods and parents' expectations of support. Both researchers used the new codes to code all the transcripts and differences were discussed and resolved. The survey data was mostly analysed using a quantitative data analysis approach, although the three free-text questions were analysed using Thematic Analysis following the same process as above.
\section{Findings}
In what follows, we first describe how our parents perceive and manage risks. We then outline children's understandings of and coping abilities with online privacy risks. Finally, we highlight parents' needs for managing their children's data privacy risks related to their children's use of mobile devices. 

\subsection{Parents' Concerns}
In both the lab study and online survey, parents said that they were generally very concerned of their children's use of these devices: only 2 families in the lab study mentioned that they never thought about privacy issues related to their children's use of mobile devices; and 44\% parents in the survey said they thought about their children's online privacy ``very often'' and 31.3\% said that they thought about it ``sometimes''. However, our online survey data also suggests that our parents' primary concerns were centred on \textit{app content}  (86.8\%), followed by the \textit{cost of the  apps} (70\%). This is consistent with parents from the lab study, as said by P1, ``\textit{before privacy, I want to know what does this do?}''.

Furthermore, despite parents' expressed concerns, our survey results suggest that many children in these families could still be using apps that were inappropriate for their age (i.e. content) or accessing their sensitive personal information. Table~\ref{tab:apps} shows children's top favorite apps that were mentioned more than 5 times in the online survey responses. For example, only 4 (out of 14) of these required no access to sensitive personal data, such as contact details on the device, location information or unique identification of the device; and five of these popular apps had an age rating inappropriate for young children aged 6-10 and would require parental guidance\footnote{The information about each app is retrieved from Google Play Store, accessed in December 2017. The UK-specific interpretation of the content rating can be found \url{https://support.google.com/googleplay/answer/6209544?hl=en-GB}. PEGI 3-7 are probably more appropriate for young children aged 6-10, than those  rated ``requiring parental guidance''.}, including YouTube as well as some popular social media apps.

The level of privacy concerns expressed by the parent participants in the survey also did not significantly correlate with the type of favorite apps used by the children: parents who were generally very concerned might still choose apps having inappropriate age rating or excessive access to their children's personal information.
%
%Neither could we find out an explanation to the choice of apps from the  online survey. However, 
In our lab study, most (10/12) parent participants expressed that they were unaware of what data about them or their children was being collected, by whom, and how it was shared with (``\textit{We have not been worried about this sharing of information before}''[P2]). Only very few parents (2/14) had gone through the privacy settings on the device prior to the study (``\textit{basically I will just check the permissions and [if] there is anything that I can turn it off then I will}''[P1]). This indicates that there is a gap between our parent participants' concerns and their actual level of awareness of technologies and information about apps.

\begin{table}[ht!]
    \centering
    \begin{tabular}{c|l|l|l}
     App & Number of use & Age rating & Sensitive personal data access \\ \hline
YouTube & 110 & Parental guidance & YES \\
Minecraft & 40 & PEGI 7 & YES \\
Roblox & 29 & Parental guidance & YES \\
Netflix & 14 & Parental guidance & YES \\
CBeebies & 10 & PEGI 3 & \textbf{NO} \\
WhatsApp & 10 & PEGI 3 & YES \\
YouTube Kids & 9 & PEGI 3 & YES \\
Clash Royale & 9 & PEGI 7 & \textbf{NO} \\
Candy Crush & 8 & PEGI 3 & \textbf{NO} \\
Facebook & 8 & Parental guidance & YES \\
Angry Birds & 6 & PEGI 3 & YES \\
Pokemon Go & 5 & PEGI 3 & YES \\
Temple Run & 5 & PEGI 3 & \textbf{NO} \\
Music.ly & 5 & Parental guidance & YES\\
    \end{tabular}
    \caption{Top Favorite Apps Mentioned in the Online Survey}
    \label{tab:apps}
\end{table}

\subsection{Parents' Approach to Risk Minimisation}
% We observed in the lab study that technical restrictions were dominantly applied to families, often combined with other mediation approaches like maintaining an ambient awareness, establishing family rules as well as passive or active mediation approaches. In literate active mediation ``refers to the process of discussing certain aspects of programs with children''~\cite{wisniewski2017parental}. 

\subsubsection{Parental restrictions}
The main strategy to enact parental restriction by parent participants in our study was the use of password control: this was widely applied by families from both the lab study and the online survey. For example, only 24\% parents reported \textit{no} password control in the online survey; and 43\% of them stated that they had controls of installation of even free apps, which seems to indicate a level of vigilance. This was consistent with our lab study participants, where all (but two) families had password controls for any app installations and device log-ins.

Furthermore, in the lab study, we also observed that a few families sought for Internet access restrictions. However, our interviews showed that parents did not necessarily fully understand the implications of their controls. For example, 2 (out of 12) families mentioned that they took out the SIM cards from the smartphones given to their children, and they only realised later that their children could still access to the Internet through wifi. 
% For example, parent of 8-year old boy (P4) realised this during the study, while describing her son's usage:  
% \begin{quote}
% ``\textit{... he is starting to use Facetime with his friends and they play `app X' together... now I realise that it does have online access; erm, now I am getting a bit more concerned about it.}''
% \end{quote}
Taken together, this suggests that restrictions might not be always effective; and even so, tend to be on the level of binary control (install/not install). The example of online connectivity restrictions highlights the challenges that parents in our study faced in fully comprehending their choice of technologies, and hence their ability to provide sufficient protection support for their children's online safety and privacy.

\subsubsection{Passive parent-led app screening}
A common pattern observed in our study was reliance on parental vetting (or direct selection) of the apps to be installed. The majority of the parents (73\%) in the online survey reported that they installed the apps for children after having screened them themselves and only 13\% thought their children might have installed the apps by themselves. 

Parents from the lab study reported a higher vigilance of app screening. The majority of the parents (10/12) in the lab study believed that their children could not install any apps without their permissions (``\textit{my mum has the password, so I can't just get it myself}'' [C5]) or their awareness (``\textit{[as they use my email address], I can see what comes in}''[P1]). Furthermore, some parents in the lab study would regularly go through the app store (``\textit{I search for free games, and categories, and then put in their ages, and type in educational}''[P5]) or external recommendations to identify suitable apps for their children and install for them. Sometimes children were involved in the process; sometimes not. 

% P2 described their app choice process for their 7-year old daughter:

% \begin{quote}
%     ``\textit{My husband and I, we looked to see what kind of games got good reviews for children and then we talked to her about what she might like... one of the sites we use is 'common sense media'...they do a review of movies and apps ....some of their staff do a review of what age they think it is appropriate but there are also user reviews underneath, they don't always agree.....so I look and see what it says in terms of the age range, and some times with the movies. }''
% \end{quote}

2 families in the lab study mentioned that to mitigate their children's risks, they minimised the number of apps installed on their devices. For example, P11 mentioned that 
\begin{quote}
``\textit{I am always very weary with apps. For me what I would really encourage are educational apps to learn for example maths and this kind of thing}''.
\end{quote}
Although these families were generally very vigilant with their children's access to apps, their primary focus remained on apps' content appropriateness and controlling their children's access. None of the above approaches would raise our parents' awareness of personal data collection by mobile apps; nor manage these risks effectively for them. 

\subsubsection{Setting home norms}
We found no consistent patterns in how in-home norms about app usage would be set or enforced by the parents. 
The majority of the parents in the lab study believed that their children would `naturally' know when to ask for help, and would require their children to seek for parents' guidance by deploying  family rules or technical restrictions. However, our interviews showed that the rules were not always consistently carried out either by the parents or children. For example, C8 described that ``\textit{I ask my mum and sometimes she allows it sometime not}''. 
Despite this belief, several families from the lab study also mentioned that they required their children to use their devices in a common family area. P7, a mother of 10-year old mentioned that ``\textit{We try to keep them downstairs. So they are not doing things we are not aware of}'', and P11 said that ``\textit{When she is on YouTube, I try to make sure I am around. To keep an eye on her}.''
%
%12/14 of our child participants used a smartphone or an iPad/Android tablet (3 of which share the device with their parents), while 2 families chose an Amazon Fire for Kids for their children in order to minimise their risks, as described by a Mum of a 6-year old boy ``\textit{one of the reasons we got the Amazon Kindle is because it is safer}''
%
These observations do not suggest that family rules are ineffective, but that executing family rules against technologies can be a challenging task. % Parents likely have the ability to adapt to different social norms in different  contexts, whereas this ability is not necessarily transferred to the children actively or consciously.

%Although this approach was not mentioned in other families, 5 families explicitly	 mentioned that they had removed some apps off their children's devices because parents noticed that the apps were displaying offensive content/promotions or asking for access to too much information. This shows that parents were generally trying to maintain an awareness of their  children's activities, and they would directly intervene when when noticed any problems. %However, none of them reported that they were doing a regular screening of apps for their threats to their children's privacy online. 
%stay downstairs
%always ask

\subsubsection{Active mediation sparingly used}
%Kumar et al~\cite{kumar2018cscw} made a distinction between an \textit{active mediation} approach, which is used by parents who are being explicitly mindful of their children's privacy risk online, and a \textit{passive mediation} approach, where discussions with children are not motivated by parents' concern of their children's privacy risks. 

In our lab study, only 2 families explicitly talked about personal data access on mobile devices with their children. For example, when being asked how he might go and install new apps on the device, the 8-year old boy responded with the following: \textit{ ...  she [my mum] is strict about privacy settings,  she has told me ...  only apps that really need it... She doesn't want our data to go to other people...they can break into your home''} [C8]. This shows how the parent (not present in the study) was being explicit about data privacy risks with the child and trying to discuss the implications of the risks as well as setting up restrictions. Both families used simple real-world threats to help children understand the possible consequences of oversharing. No additional technologies or online resources were mentioned.

%A parent from another family with a 10-year old boy described how her family recognised the importance of talking about privacy and security issues from a young age: ``\textit{I'd like to think that they just know what to ask. [We started to] talk about permissions from the very beginning}.''[P1]
%\comm{how is this quote related to the rest of the section .. are these two the only families that talked about personal data access? If so, wouldn't it be more interesting to talk about what the others did not say? Or said instead? Feels weird to highlight two (random) good examples in a see of bad ones(?)}
We observed two themes in the rest of the families from the lab study, who did not take any active mediation approaches yet at home. One theme was that some parents believed that privacy was a concept too hard for young children and they would avoid active mediations and take a parent-led protective approach.
For example, the mother of a 6-year old girl concluded that 
\begin{quote}
``\textit{... she can't understand. .. it's my responsibility and it's hard to expect a child to understand that these companies are actually manipulating their naivety}''[P9]
\end{quote}

% Another mother of a 6-year old boy reflected that ``\textit{he has no understanding of what it means to give access to pictures. it's solely dependent on the parent being wise enough about it.}''. 	

The other theme was that a few parents were open to introducing early discussions with their children, but they struggled with communicating to their children, for example, P11 described that ``\textit{This mostly comes from the school. When it comes from the parents, they don't really [listen].}'', or felt ill-equipped with their own understandings ``\textit{I feel [the other parent] is more equipped to have these conversations. I barely know how to install an app.}''[P5].

This suggests that the importance of actively mediating their children's awareness from a young age is not yet widely recognised by parents in our study, and these parents showed their struggle with keeping up with the technologies or finding a way to discuss these issues with their children effectively.

\subsection{Children's Awareness of Personal Data Privacy Threats}
Most parents in the online survey were not concerned about their child's awareness of online privacy risks: 83.3\% parents reported that they were ``very'' or ``somewhat'' happy, and a high proportion of them (77.7\%) expected that their children would ask for help or click 'no' to unnecessary requests for personal information access by the apps.

However, these patterns were not supported in the lab study. Despite indications of emerging understanding about sensitive information by our child participants, and good coping strategies from older children (above 8 years old) to directly experience-able risks, like in-app promotions or inappropriate content, children largely demonstrated a knowledge gap regarding personal data privacy threats related to their use of mobile devices and younger children particularly struggled.

Although most of our child participants (9/14) demonstrated recognition of sensitive types of information for them, including pictures, location information, passwords, email addresses, medical records, bank details etc, the main threats perceived by them resulting from exposing their personal information would come from \textit{strangers}. 

% For example, an 10-year old boy ([C9]) expressed that 
% \begin{quote}
% ``\textit{I don't like anyone else to have access to my pictures... I once saw it on a TV adverts, which says what might happen if you share a picture with you standing in front your house.}''
% \end{quote}

%\subsubsection{Recognising some risks but not primarily for privacy concerns}

The majority  of the children (12/14) in the lab study mentioned that they could notice risks like \textit{in-app promotions}, \textit{inappropriate content}, or \textit{requests to sensitive information like credit card details or passwords}; and they used approaches like \textit{skipping/declining access} or \textit{asking for help from parents}. However, when an advert or promotion was skipped, the main reason was mainly about being bored or annoyed ("\textit{I don't really want to watch. because I want to watch something else}''[C10]). When help was sought from parents, it was more likely to be due to insufficient access to the device or a vague feeling of something not right (``...\textit{I didn't really like the look of, so I go to my Daddy}..''  [C2]). This suggests that our children may not always fully comprehend the actual risks or recognise the privacy-invasive context.

%\subsubsection{Struggling with personal data access by apps}
When we asked the child participants what they thought about one of their favorite apps requesting access to their device's camera, location information or photos,
most of the children couldn't answer, or gave unintelligible replies, such as ``\textit{I don't know what would happen. I will press yes}''[C11: a 6-yo boy].
Only a few older children (4/14) demonstrated some abilities of recognising potential risks, and associate specific social norms for different contents, like different apps, different types of personal data, or different data recipients. For example, C8, an 8-year old boy, when being asked why he thought location information was sensitive, he mentioned that ``\textit{only apps that really need it, like maps and stuff  ... you won't even let the app that tracks weather and stuff to have location [because] they can break into your home}''.

% C9, a 10-year old boy said that ``\textit{I think I wouldn't want to do that. I wouldn't want to share this kind of information. i can share simple things, like camera. but i think location very sensitive}''.

% C13, an 8-year old girl, explicitly mentioned the possible information recipients of her data, by saying that ``\textit{I never let them. Because it might be a complete stranger and I don't trust people like that.}''

%Although these children's prior knowledge did not explicitly mention personal data access by mobile apps, they demonstrated an ability of applying their knowledge to these new contexts. 

\subsection{Children's Coping Strategies} 
\subsubsection{Relying on parents' help}
Parents' guidance was widely sought by our child participants to the lab study. However, we observed two themes in children's reliance on parents: 1) a few children were consciously recognising potential risks and followed strict family norms, while 2) most children were generally motivated by a vague feeling about the situations or were restricted by what they could do on the devices by themselves. 

For example, a 10-year old boy [C1] recognised the risk of an app asking for access to his audio and described that ``\textit{I do know when to ask...}'' and C8, an 8-year old boy, responded to our question about ``\textit{how he would respond to an app requesting for access to his camera}'', by saying that ``\textit{i ask my mum ... because she is strict about privacy settings}''. The majority of the children reacted passively to the risks, for example, C7 mentioned that ``\textit{Dad normally test whether these things send photos}''[C7]), and she would rely on Mummy to safeguard her privacy settings. 

This implies a demand on parents' technical competence and also a need to raise children's ability of recognising risky situations. It is hard to be thorough with parental controls, given the complexity of technologies. Helping children, who are in the frontier of these risks, to recognise and flag risks, is a crucial piece in safeguarding children's privacy online.

\subsubsection{Declining access}
Our child participants' ability of declining personal data access request by mobile apps correlates to their awareness about personal data collection practices of the apps.
The majority of the children in the lab study showed a reasonably good ability of declining in-app game promotions or adverts. However, they were largely unable to make a conscious decision with apps asking for access to cameras or personal details. Most children had not been told about what they can/should do. If there were a ``cross''  button and they were annoyed and wanted to get on with their games, they would just tick it away. 
Only a few children demonstrated some more developed mental model regarding their decisions, which took into consideration of the needs or purposes of an app asking for access to particular information. For example, an 8-year old boy described that  ``\textit{I don't want `chrome' to have access to my phones... so it can show to others. and 'stickbot' doesn't really [need] either}'' [C5]. 

This indicates a critical gap in our child participants' awareness and coping skills regarding personal data collection by mobile apps. This is contributed by parent participants' own lack of awareness of the risks but also their lack of engagement with the children, believing that they are too young to understand these issues.

% C8 elaborated in his response that ``\textit{we don't know what it [an alarm clock app] would need it [location info] for}''.

% C13, a 7-year old girl described her thought process when decling an app asking for her personal information by saying that ``\textit{I did not know exactly know what they were going to make me do ... Because in my school they said you should never go on to something that isn't very good to go on. So, I just use my mind. At school they said. Because sometimes people could learn things about you. If you type in things about your life and thing your home address come contact you.}''. 

\subsection{Parents' Reflections on Personal Data Collection}
\subsubsection{Technologies hard to follow}
In the survey, parents expressed a high level of concerns about apps accessing their children's location information (93\%) or cameras (64\%). However, less than 20\% of the parents checked their children's privacy permission settings regularly or when they installed an app. On the conclusion of the survey, nearly 86\% of them felt that they learnt something through the survey and would need to go and check their children's apps and privacy settings more carefully. This suggests that parents' needs are largely unmet, and there is a critical gap between what parents are concerned about and what they are supported with. 

In our lab study 7 parent participants expressed that they found it hard to understand technologies and keep up to speed. They often relied on the other partner, who may be a Dad or Mum, to take a lead on safeguarding the children. For example, P2, a mother of a 7-year old girl,  expressed at the conclusion of the interview that ``\textit{he [my husband] understands more than I do.  I don't know how to do that. We keep on discussing...at some point you need to show me how to do that.}''.  

Four parents thought they had been careful with the device controls by applying restricted access to Internet or content. However, once knowing more about the personal data privacy threats, they found it hard to be thorough,  ``\textit{I thought I was in a good control of what they have access to, and what they have told me. We ought to be more sensitive}'' [P4].

\subsubsection{Feeling losing control}

At the conclusion of the lab interviews, the majority of the parents (9/12) expressed an increase of concerns and their surprises of not being made known. They demanded for \textit{more transparency} in this space (``\textit{when you choose an app very clearly just one page just say this is what we do with your information. this is where this is going, rather than some kind of hidden secret}''[P10]) or legal regulations (``\textit{Legally ... we should be protected by law. They [these companies] should simply and very clearly show all these things that they are doing}''[P3]).

A few parents in the lab study reacted more passively to the issue, expressing that they believed that these issues should be dealt with by relevant authorities in due course (``\textit{I am trusting other people do know, and exercising the influence and pressure}'' [P5]) or felt that this type of personal data collections was unsurprising and hard to avoid given how the current app ecosystem works (``\textit{There is nothing to stop trackers, it's just a polite thing that you subscribe to a mailing list, but they don't have a way to unsubscribe me}''[P1]). 

% \subsubsection{Passive technical controls}
% % parents decide
% Almost all our participant families have set up some parental controls on the devices, including restrictions on installation of apps, access to Internet, or time limits for the device usage. Most family devices were linked to parents' email addresses, so that parents maintained an awareness and kept a control of what children were able to install or access to. No parents in our lab study reported using any blocking, content filtering, or monitoring software at home or on the devices that children used. However, as mentioned previously, most of these restrictions were used to control the content children could access, their screen time, or prevent them from being approached by strangers online, instead of out of concerns of their personal privacy.

\section{Discussions}
% a recap of study findings
We found that child participants in our lab study demonstrated an awareness of risks that were directly experience-able, such as in-app promotions or inappropriate content. However, they had limited knowledge about any of the more insidious personal data privacy threats and they currently largely relied on parents' help to safeguard their online safety or provide guidance. These risks were neither fully recognised by their parents nor regularly discussed with them. 

Our lab study and survey results further showed that our parent participants played a key role in mediating their children's interactions with mobile technologies. However, our analysis of both the lab interviews and the online survey showed that our parent participants largely had limited understanding about personal data collection by mobile apps, and their current concerns largely centred around app content, in-app promotions and exposure to strangers online. Parent participants in both studies largely took a passive approach to safeguarding their children, by setting up technical barriers or strict family rules. Our interviews showed possible reasons behind our parent participants' mediation approaches, including their perception that their children were too young to understand privacy risks, and their struggle of communicating with their children effectively. Existing literature has suggested that it is critical to consider active parental mediation of children's technology usage from an early age~\cite{hashish2014involving,kumar2018cscw}, when they are still interested in developing trusty  relationship with grown-ups. This suggests that there might be a critical gap in exiting technologies and resources for not only raising parents' privacy awareness, but also their recognition of the importance of combining technical controls with active social interaction and mediation. 

\subsection{Helping Parents to Raise Their Privacy Awareness}

Our parent participants' current concerns about their children's usage of mobile devices in terms of privacy risks are largely misplaced. Our data suggest that this is partially due to the difficulty parents expressed about keeping up with rapid developing technologies, and the lack of available tools that would be  focused on raising users' awareness of privacy risks. 

Privacy researchers have explored various ways to enhance the current opaque and passive privacy `permission' models on mobile devices, by contextualising data access purposes~\cite{shih2015privacy}, summarising risks~\cite{lin2012expectation}, augmenting permission interfaces~\cite{liccardi2014no},  and exploring just-in-time notification/nudging~\cite{almuhimedi2015your,balebako2013little}. However, in our study, we still found parents in our lab study struggling with having to scroll down to the bottom of the app page to read a piece of lengthy and jargon-laden  privacy policy as their only option. This suggests that these technologies are not yet becoming accessible in the mainstream platforms. 

Communicating privacy risks---especially those without immediate impacts---to busy parents can be challenging. As we see, even though our parent participants in our survey largely focused on the importance of content appropriateness, their young children were still exposed to apps that were probably inappropriate to their age (see Table~\ref{tab:apps}). The age rating of apps can be obscure (depending on the device and platform), and recognising apps that are indeed designed for young children in the general app store can be challenging even for trained eyes. This suggests that designing for parents  may require careful consideration about information accessibility and usability. A summarised presentation of app risks~\cite{lin2012expectation} can provide a convenient shortcut for parents to quickly sift through apps with less risks. However, privacy risks can be a personal and contextualised decision making process~\cite{nielsen2002getting}, and a one-fit-all  might not work for everyone's need.

Another challenge is that parents can face a more complicated social context, such as the peer pressure their children feel from their friends or siblings~\cite{zhang2016nosy,wisniewski2017parents}. These unique social factors could lead to either a more complex mental model or compromises to be made. Designing for parents thus needs to consider additional social context related to the technologies, such as experience comments from peers or other parents. They may  provide as valuable inputs to their decision making as knowing 
the specific companies or jurisdictions that have access to their children's personal data.

\subsection{Filling in the Parental Mediation Void}
Our study not only highlighted the knowledge gap in parent participants' awareness of personal data privacy threats, but also the even bigger knowledge gap of their young children's. This is alarming, given that our results showed that our child participants were already exposed to the risks, and they were unclear when to ask for help from their parents. In our study, our child participants showed a strong reliance on their parents' help, and a strong and trusty relationship with their parents. This indicates that there might be a great opportunity for parents to take a more proactive role in facilitating children's understanding about privacy risks.

However, currently our parent participants largely believed either that their children were too young to understand or they found themselves ill-equipped to mediate their children. This suggests there might be a critical gap in modern parents' awareness of the importance of active mediation and in technologies supporting parents to take an interactive approach.
This is analogous to the situation of parenting approaches to teenage online habits. A recent review of parental control tools~\cite{wisniewski2017parental} has shown that current technologies largely focus on parental restrictions/monitoring for parents of teenagers, and provide limited support for increasing parents' awareness or facilitating parent-child discussions, which is in fact desired by both parents and their teens. Existing studies have demonstrated that a co-learning process can facilitate a more effective mediation of technology usage between parents and young children~\cite{hiniker2016screen,hashish2014involving}

Our findings suggest the need to provide more support in raising parents' awareness on the importance of discussing these issues with their children from a young age. Existing resources online are largely focused on older children (like tweens or teenagers) and do not support younger children or cover this type of emerging personal data privacy risks. Although apps or smart toys targeted for young children are currently in the limelight due to rumours about them being unsafe and prompting children to reveal personal information~\cite{anglea2014,mcreynolds2017toys}, understanding of the risks associated with smart apps or smart toys is still scarce, and parents are only provided with generic cautions instead of specific advices.

Technologies should also consider incentivising parents to actively discuss these issues with their young children by providing specific age options, and content specially targeted for a younger age group. Literature has suggested that parents of teenagers often face stronger push-back from their older children for interfering or mediating their use of technologies~\cite{wisniewski2017parents}. Kumar et al~\cite{kumar2018cscw} argue that parents could miss a critical opportunity to establish a mediation relationship with their young children if they choose to delay the discussions, because ``[young] children are more focused on adult/parent relationships''. 

\subsection{Scaffolding Children's Understanding}

The biggest challenges faced by our child participants in the lab study were that they found it hard to understand how information was used by mobile apps, who could access information and where information could go, and how to react to different risks, by recognising when to ask for help. This can be regarded as a complex topic for young children, and the ICT curriculum in the UK is still forming and going through rapid changes in the last few years. However, we argue that children's early understandings could have potential long-term impact on their mental model when they grow older and have more freedom with their smartphone devices~\cite{kumar2018cscw}. 

The Zone of Proximal Development (ZPD) theory from Education Psychology~\cite{chaiklin2003zone} suggests that children's learning is better supported if education starts with individual child's current zone of knowledge and supports their learning process according to their individual needs and environment. A recent report in the UK on Digital Childhood also highlighted a gradual change of focus is needed for mediating young children's use of digital technologies according to their developmental stages~\cite{digital2017}. The critical knowledge gaps identified in our study could provide initial indicators to future privacy education technologies. 

%Our study shows that younger children face an even bigger knowledge gap and their parents would largely take a a protective approach. This is alarming, given that our study shows that young children are already exposed to potential privacy risks, and they are not mediated to learn and they could miss the recognition of risks and expose themselves.  

% Older children have also demonstrated an ability of applying their existing privacy knowledge to newer, less familiar contexts. This suggests that education could facilitate young children's understanding of this seemingly abstract concept from an early age. Extending existing literature, we identified critical gaps in children's understandings about information sharing on mobile platforms and its implications for their personal privacy. 

% \subsubsection{Strengthening children's understanding about information transmission on mobile platforms}

%Not only young children but also parents appeared to have difficulty to recognise or fully understand how information transmits on the mobile platforms. Several families mentioned that they ``removed SIM cards'' in their children's smartphones in order to reduce risks without realising that children could still have full access to Internet. A few parents were also surprised that their restrictions to Internet access were actually not as effective as they thought. 

Furthermore, the web prototype we used in the last part of our lab study provided some possible future design directions for us to  explore. While walking through children's favorite apps in the prototype (see Figure~\ref{fig:tracker}), we observed that older children (aged 8-10) found it ``very interesting''. They demonstrated an understanding of the information transmissions through interactions with the map view, and engaged in the discussions about the companies and countries to which their favorite apps might be sending their personal information. They responded that they believed that ``\textit{it's not right for companies from another country to access my information}'' [C9, aged 10] or that ``\textit{[I am ] sort of surprised, but not much. because people would do anything to make more money these days}'' [C1, aged 10]. %However, these children probably did not fully grasp the implications of these data collection behaviours. For example, although an 8-year old boy was very surprised by the information tracking by expressing ``\textit{why did i not know that?}'', he also thought that ``\textit{it might actually be useful if they advertise stuff that you will find interesting and useful}''.
% * <rdpbinns@gmail.com> 2018-01-17T16:37:50.401Z:
% 
% > \textit{it might actually be useful if they advertise stuff that you will find interesting and useful}''.
% I would say it's arguable that this boy DOES grasp the implications of these data collection behaviours; he just happens to agree with the advertising industry!
% 
% ^.
At the same time we observed that the application enabled an engagement between parents and their children.  Parents found it a useful tool to explain to their children the implications (``\textit{now you think that these companies are kind of using these as clever tools to gather your data?
you know it [means] that it's more than a game, did you?}''[P4]). This suggests that a child-friendly visusalisation of the information transmissions behind mobile apps could have the potential to not only increase children and parents' understanding of personal data privacy risks but also provide an interactive tool for co-learning.

\subsection{Limitations}

The primary limitation of this study is our sample population; our lab study participants were all residents of the affluent South East area in the UK and highly educated. We fully acknowledge that this is not representative of the populations. The complementation of this interview data with an online survey that incorporated a wider and more diverse population was one step towards validating the main themes. Furthermore, we have also participated in a large-scale science outreach activity for young children in the Autumn of 2017, with $>$100 families participating on the day. We observed similar reactions from the children, who came from a wider part of the area. %We intend to follow up the study by removing the constraints of running the interviews in the university premises, to approach schools and families from more diverse backgrounds and circumstances. 
% * <rdpbinns@gmail.com> 2018-01-17T17:36:06.445Z:
% 
% > We observed similar reactions from the children, who came from a wider part of the area.
% Not sure we can really rely on this to mitigate the problem, since we didn't formally study them, nor do we know their backgrounds
% 
% ^.

The second limitation is the lab study setting, where the presence of parents and children in the same room might influence the responses from the child participants. Furthermore, the unfamiliar setting and time constraints of the study might also limit their ability of reflective thinking. We plan to explore focus group studies with young children in future research, to reduce the influence from the interviewers and parents, and observe them in their familiar settings, like schools. Through this we aim to achieve a deeper understanding about the familiar terms used by the young children to describe risks and the building blocks required for them to adapt their existing knowledge to understanding new kinds of information transmission and associated risks.

%\comm{add here that we're aware that what children and their parents tell us they think about privacy in the special setting of coming into the lab - or even what they report in the survey - is not necessarily identical to what they "really" think or do. Refer to Social desirability bias plus the privacy paradox.}

\section{Conclusions}

In this report we presented the results of a mixed-method study towards understanding what data privacy threats children ages 6-10 and their parents were aware of during the interaction with tablet computers. We found that although our child participants had a reasonable awareness of familiar risks, like in-app promotions or inappropriate content, they (particularly younger children) largely struggled to understand the information transmissions of mobile apps, and how their personal data might be collected and pose threats to their personal privacy. Older children in our study demonstrated some strategies of declining access to information, however, our child participants still mostly relied on parents' help to mitigate risks.

Our parents also showed limited understanding regarding what data was collected by the apps and shared with whom. Their primary privacy concerns were about content and screen time controls. To mitigate their concerns, they sought passive technical restrictions, app limitations, or external information sources. However, these tools and resources largely focused on content control and provided no support for raising parents' awareness or facilitating discussions with their children. This is therefore a critical gap in the current landscape.

Based on these findings, we recommend 1) raising the general awareness of considering scaffolding children's knowledge about data privacy from an early age, where they are still at an age of seeking for a positive relationship with grown-ups, and are already facing risks at a regular basis; and 2) tool and resource developments that focus on facilitating skill and knowledge building for both parents and their young children, and encouraging an active co-learning experience. Our identification of children's key knowledge gaps and feedback to an exiting prototype could provide potential inputs for the design of future privacy education technologies.

\section{Selection and Participation of Children}
In our ethics approved study, children were recruited through public invitations shared with parents on social media, posters in permitted public spaces, mailing lists, and snowball sampling. The study took place over school summer holiday. Thus, no recruitment took place through schools. Participants lived in the South East area of the U.K. Children were limited to the ages of between six and ten inclusively. All children from each family were encouraged to participate in the study as long as they were within the age limit and did not express distress. Parents signed an informed consent and children signed an assent form with their parents' help (if needed). The forms explicitly requested consent for the study to be audio-recorded for the purposes of transcription.

\bibliographystyle{SIGCHI-Reference-Format}
\bibliography{kids-studies,chi-2017,toit-2017}

\end{document}